\documentclass[twocolumn,showpacs,preprintnumbers,amsmath,amssymb]{revtex4}
\usepackage{graphicx}
\usepackage{dcolumn}
\usepackage{bm}

\newcommand{\be}{\begin{equation}}
\newcommand{\ee}{\end{equation}}
\newcommand{\bea}{\begin{eqnarray}}
\newcommand{\eea}{\end{eqnarray}}
\newcommand{\mbb}{\mathbb}
\newcommand{\ti}{\times}
\newcommand{\half}{\frac{1}{2}}

\newcommand{\mc}{\mathcal}

\begin{document}
\preprint{OUTP-09/17P}
\title{On Gauge Threshold Corrections for Local IIB/F-theory GUTs}
\author{Joseph P. Conlon, Eran Palti}
\affiliation{%
Rudolf Peierls Centre for Theoretical Physics, 1 Keble Road, Oxford, OX1 3NP, UK}%
\begin{abstract}
We study gauge threshold corrections for local GUT models in IIB/F-theory. Consistency with
 holomorphy requirements of supergravity and the Kaplunovsky-Louis formula implies that the unification scale is enhanced by the bulk radius $R$ from 
the string scale to $M_X =RM_S$. We argue that the stringy interpretation of this is
 via a locally uncancelled tadpole sourced by the hypercharge flux.
 This sources closed string modes propagating into the bulk; equivalently open string gauge coupling running up to the winding scale $M_X$. The 
enhancement to $R M_s$ is tied to GUT breaking by a globally trivial hypercharge flux and will occur in all models realising this mechanism.
\end{abstract}
\pacs{11.25.Wx, 12.10.Kt, 11.25.Mj}
\maketitle

\section{Introduction}

The apparent unification of gauge couplings at an energy scale $M_{GUT} \sim 2 \ti 10^{16} \hbox{GeV}$
is one of the most intriguing results in theoretical particle physics. The combination of this
unification and the organisation of much of the Standard Model into GUT multiplets
suggests a more fundamental structure involving a unified gauge group $SU(5)$, $SO(10)$ or $E_6$
broken at a very high energy scale. The proximity of $M_{GUT}$ to $M_P$ also suggests a relationship with
string theory or whatever physics is responsible for quantum gravity.

However it is unclear whether new physics arises precisely at $M_{GUT}$.
The precise relationship between GUT, Planck and string scales is determined by
threshold effects from loops of heavy particles.
The actual masses of such heavy states in relation to $M_{GUT}$ is important for various
phenomena, such as the rate of proton decay induced by the heavy $(\bf{3}, \bf{2})$ gauge bosons completing
the $\bf{24}$ $SU(5)$ multiplet or the scale of neutrino masses induced by
heavy right-handed neutrinos.
Given a proposed GUT model it is therefore important to compute the threshold effects that affect the
interpretation of the scale $M_{GUT}$.

In string theory
the traditional setting for GUT model building has been the heterotic string with Wilson line breaking.
However recently techniques have been developed for GUT model building in the context of F-theory/IIB \cite{08022969, Beasley:2008dc,  
Blumenhagen:2008zz}.
Here the GUT is realised on a stack of branes wrapping a shrinkable 4-cycle (i.e. a del Pezzo surface). In the IIB context
only $SU(5)$ GUTs can be realised but in an F-theory context also $SO(10)$ or $E_6$ gauge groups are possible, as well
as perturbatively forbidden couplings such as ${\bf 10} \, {\bf 10} \, {\bf 5}$.
 Chiral matter lives on 2-cycles at the intersection of the GUT brane with $U(1)$ branes.
Such a geometry also fits well with the LARGE volume scenario for moduli stabilisation for IIB orientifolds, where
full moduli stabilisation and low scale supersymmetry breaking requires the Standard Model to be realised on branes wrapping shrinkable 4-cycles 
\cite{hepth0502058}.

One characteristic of such F-theory GUTs is that they are local.
Recently detailed analyses of the form of threshold corrections for local models have been carried out in \cite{Conlon:2009xf, Conlon:2009kt}
for D-branes at orbifold/orientifold singularities. Such models have universal
tree level holomorphic gauge coupling. However the 1-loop running is non-universal and splits the gauge couplings.
In orbifold language the loop corrections can be attributed  either to $\mc{N}=1$ sectors (associated to fully twisted closed string modes)
or $\mc{N}=2$ sectors (partially twisted modes).
It was shown in \cite{Conlon:2009xf, Conlon:2009kt} that running consists of two phases. $\mc{N}=1$ sectors run from the string scale while
$\mc{N}=2$ sectors run from the enhanced scale $RM_s$, where $R$ is the bulk radius.

In this paper we extend this analysis to the phenomenologically interesting case of local GUTs with $U(1)_Y$
hypercharge breaking. Such models reside
in the geometric regime away from the orbifold point.
For concreteness we work in the IIB setting but all results should carry over to F-theory.

\section{Local GUTs and field theory}

The GUT brane wraps a del Pezzo divisor $D$ with matter realised on the intersection of $D$ with other $U(1)$ branes.
At tree-level the $SU(5)$ universal gauge kinetic function is given by
\be
f_{SU(5)} = T = \frac{1}{2g_s}\int_{D} J \wedge J + i\int_{D} C_4 \;.
\ee
Here $J$ is the Kahler form with the volume of $D$ given
by ${\cal V}_{D}=\frac12 \int_{D} J \wedge J$. $C_4$ is the RR four-form and $g_s$ is the string coupling $e^{\varphi}$.

GUT breaking is accomplished by turning on an internal world-volume hypercharged flux $\mc{F}_Y$
on a 2-cycle of the del Pezzo $\mc{C} \in H^2(D, \mbb{Z})$.
The flux $\mc{F}_Y$ also splits the Higgs doublets by piercing the 2-cycle on which the $\bf{5_H} + \bar{\bf{5}}_H$
GUT Higgs multiplet is supported \cite{08022969, Beasley:2008dc}.
There is a danger that $U(1)_Y$ could become massive via the potential Green-Schwarz coupling
\be
\int_{\mbb{M}^4} \tilde{C}^{\alpha}_2 \wedge F_{U(1)_Y} \int_{D} {\cal F}_Y \wedge i^*\left( \omega_{\alpha} \right)    \;. \label{gsterm}
\ee
Here we denote the 4D space  by $\mbb{M}_4$ and decompose the RR four-form $C_4 = \tilde{C}_2^{\alpha} \wedge \omega_{\alpha}$, with $\omega_{\alpha}$ 
being a basis of two-cycles $\omega_{\alpha} \in H^2(\mc{M}_6)$. However provided
${\cal F}_Y$ is locally non-trivial but globally trivial then
\be
\label{GS}
\int_{D} {\cal F}_Y \wedge i^*\left(\omega_{\alpha}\right) = 0 \;, \;\; \forall\; \omega_{\alpha} \in H^2\left({\cal M}_6\right) \;.
\ee
The Green-Schwarz term is therefore absent and hypercharge remains massless \cite{0610007}.

The GUT branes contain two relevant $U(1)$ bundles: $\mc{F}_D$ valued in the diagonal $U(1)$ of $U(5)$ and
the hypercharge bundle $\mc{F}_Y$ valued in $U(1)_Y$ (note $\mc{F}_D$ may also contain parts trivial in the Calabi-Yau).
The tree-level holomorphic gauge couplings for the $SU(3)$, $SU(2)$ and $U(1)_Y$ factors are given by \cite{08022969, 08120248}
\bea
\label{couplings}
f_{SU(3)} & = & T - \half S \int_{D} \mc{F}_D^2 \equiv T + \alpha_3 S, \\
f_{SU(2)} & = & T - \half S \int_{D} \left[ \mc{F}_D^2 + \mc{F}_Y^2 + 2 \mc{F}_D \mc{F}_Y \right] \equiv T + \alpha_2 S, \nonumber \\
\frac{3}{5}f_{U(1)_Y} & = & T - \half S \int_{D}  \left[  \mc{F}_D^2 + \frac{3}{5} \left( \mc{F}_Y^2 + 2 \mc{F}_D \mc{F}_Y \right) \right] \nonumber \\
& \equiv & T + 
\alpha_Y S.
\nonumber
\eea
Here $S$ is the axio-dilaton superfield $S=e^{-\varphi} + iC_0$. There are additional one-loop corrections to $f_a$ depending on complex structure
moduli (see \cite{Donagi:2008kj} for a thorough study), so that overall
\be
f_a = T + h_a(F)S + \epsilon_a(U),
\ee
where $a$ is the gauge group index.
Any extra perturbative dependence on $T$ (or other K\"ahler moduli) is forbidden by the combination of holomorphy of $f_a$ and the shift symmetry
$T \to T + 2 \pi i$.
For the notion of unified gauge couplings to make sense we require that $h_a(F)$ (if non-universal)
and $\epsilon_a(U)$ are much smaller than $T$.
Note that $h_a(F)$ are in fact universal if both ${\cal F}_Y$ and ${\cal F}_D$ are roots of $E_n$ with
${\cal F}_Y \cdot {\cal F}_D = 1$ \cite{08120248}.

However the study of gauge coupling unification requires use of physical rather than holomorphic couplings, with the two related by
anomalous rescalings.
In general, physical and holomorphic gauge couplings are related by the Kaplunovsky-Louis formula \cite{9303040, 9402005},
\begin{widetext}
\bea
\label{KL}
g_a^{-2}(\Phi, \bar{\Phi}, \mu) & = & \hbox{Re}(f_a(\Phi)) +
 \frac{\left( \sum_r n_r T_a(r) - 3T_a(G)\right)}{8 \pi^2}
\ln \left( \frac{M_P}{\mu}\right) + \frac{T(G)}{8 \pi^2} \ln g_a^{-2} (\Phi, \bar{\Phi}, \mu) \nonumber \\
& & + \frac{(\sum_r n_r T_a(r) - T(G))}{16 \pi^2} \hat{K}(\Phi, \bar{\Phi})
 - \sum_r \frac{T_a(r)}{8 \pi^2} \ln \det Z^r(\Phi, \bar{\Phi}, \mu).
\eea
\end{widetext}
Here $g_a^{-2}(\Phi, \bar{\Phi}, \mu)$ is the physical coupling, $f_a(\Phi)$ the holomorphic coupling, $\mu$ the energy 
scale, and $\Phi$
light uncharged moduli superfields. $\hat{K}$ is the moduli K\"ahler potential and $Z^r$ are the matter field
kinetic terms.

In the large volume limit where gravity is `decoupled', the anomalous terms in (\ref{KL}) receive large logarithmic enhancements by the bulk volume
$\mc{V}$.
In IIB orientifolds, $\hat{K} = -2 \ln \mc{V}$ \cite{0403067} while locality of the physical Yukawa couplings requires
that $Z_{\alpha} \sim \mc{V}^{-2/3}$ \cite{Conlon:2006tj}. This follows from the supergravity expression for physical
Yukawa couplings,
\be
\hat{Y}_{\alpha \beta \gamma} = \frac{e^{\hat{K}/2} Y_{\alpha \beta \gamma}}{\sqrt{Z_{\alpha} Z_{\beta} Z_{\gamma}}} \;,
\ee
and the fact that holomorphy and shift symmetries implies that $Y_{\alpha \beta \gamma}$ cannot depend on the K\"ahler moduli
$T_{\alpha}$. Locality is the statement that physical Yukawa couplings do not depend on $\mc{V}$ and (assuming different local
fields see the bulk volume in the same way) this gives $\hat{Z}_{\alpha} \sim \mc{V}^{-2/3}$.

Using these relations in equation (\ref{KL}) gives
\be
\label{mirage}
g_a^{-2}(\mu) -\frac{T(G)}{8 \pi^2} \ln g_a^{-2}(\mu) =\hbox{Re}(f_a(\Phi)) + \beta_a \ln \left( \frac{(RM_s)^2}{\mu^2} \right),
\ee
where we use $M_s = M_P/\sqrt{\mc{V}}$ and $\beta_a$ are the field theory $\beta$ functions determined by the massless modes.
The effective energy scale from which the physical couplings (\ref{couplings}) appear to run is therefore given by ${\cal V}^{\frac16}M_s \equiv R 
M_s$ rather than
$M_s$: the unification scale is, up to a model-dependent numerical prefactor, the winding mode scale rather than the string scale.
The size of this prefactor cannot be determined; however its effects will be diluted in the limit $R \to \infty$.
Note the volume $\mc{V}$ cannot appear in $f_a(\Phi)$
due to the constraints of holomorphy.

At a microscopic level threshold corrections come from heavy string/Kaluza-Klein modes which appear for example
as Higgs triplets or $X/Y$ gauge bosons. Away from the orbifold point it is not possible to compute the microscopic spectrum
explicitly. However in a supersymmetric effective theory the effects of all heavy modes can be parametrised by corrections to the
effective action. From (\ref{KL}) 1-loop corrections to $\hat{K}$ or $Z^r$ correspond to 2-loop corrections to gauge coupling running
and can be ignored at our level of approximation. Therefore the only effect of
heavy Kaluza-Klein modes, heavy Higgs fields and heavy gauge fields in (\ref{KL}) is to contribute to
the 1-loop holomorphic correction to the gauge kinetic 
function. However holomorphy then implies that the magnitude of such a correction is independent of the bulk volume.

There is a potential subtlety regarding the holomorphy of the gauge kinetic function.
The gauge kinetic function has to be holomorphic with respect to chiral superfields.
However the definition of the chiral superfields may be altered at one loop.
This comes through the presence of linear multiplets, which are common in string
compactifications and which need to be dualised to chiral multiplets. This dualisation procedure receives a 1-loop correction, which is
accounted for by a 1-loop redefinition of the chiral superfields
\footnote{The redefinition can correct the Kahler potential and we expect that the Kahler potential for $T$ 
receives a 1-loop correction such that $T+\bar{T} \rightarrow T+\bar{T} + \alpha \mathrm{ln}{\cal V}$, with $\alpha$ a constant. 
However since $T$  has not been set via an explicit moduli stabilisation procedure it is essentially chosen to match 
observations and so is unaffected by this.}. However as $T$ enters the holomorphic gauge
couplings $f_a(\Phi)$ in an $SU(5)$ universal fashion any redefinition does not affect the scale of
gauge coupling unification. This is a consequence of the GUT structure -
for branes at orbifold/orientifold singularities the couplings of twisted
 blow-up moduli are non-universal and the 1-loop redefinition breaks gauge universality \cite{Conlon:2009kt}.
Note that the dilaton and complex-structure superfields always appear as chiral multiplets and so do not receive this particular 1-loop redefinition.

To summarise, we have argued, using the holomorphy of the gauge kinetic function and an understanding of any non-holomorphic field redefinitions, that 
for local IIB/F-theory GUTs the Kaplunovsky-Louis formula (\ref{mirage}) implies that the unification scale is enhanced from the string scale to the 
winding scale. For constructions with a large bulk volume this represents a substantial effect.

\section{The string picture}

While the above field theory argument is robust and ultimately follows from the consequences of holomorpy and locality, the origin of
the scale $R M_s$ is obscure.
In \cite{Conlon:2009xf, Conlon:2009kt} the correctness of this scale was checked by explicit string calculations.
Away from the orbifold point such an exact 
calculation is a far more difficult prospect. However the physics found in \cite{Conlon:2009xf, Conlon:2009kt} easily extends to the
geometric regime and so we
now use the geometry of local GUTs to explain how the scale $RM_s$ arises from the string perspective.

In \cite{Conlon:2009xf, Conlon:2009kt} the key point underlying the enhancement to $R M_s$ is the well-known equivalence between
open string 1-loop diagrams and closed string tree diagrams (see figure 1). In particular, ultraviolet finiteness of
threshold corrections to gauge couplings is equivalent in closed string channel to tadpole cancellation for
closed string RR fields.
\begin{figure}
\includegraphics[width=7cm,height=5cm]{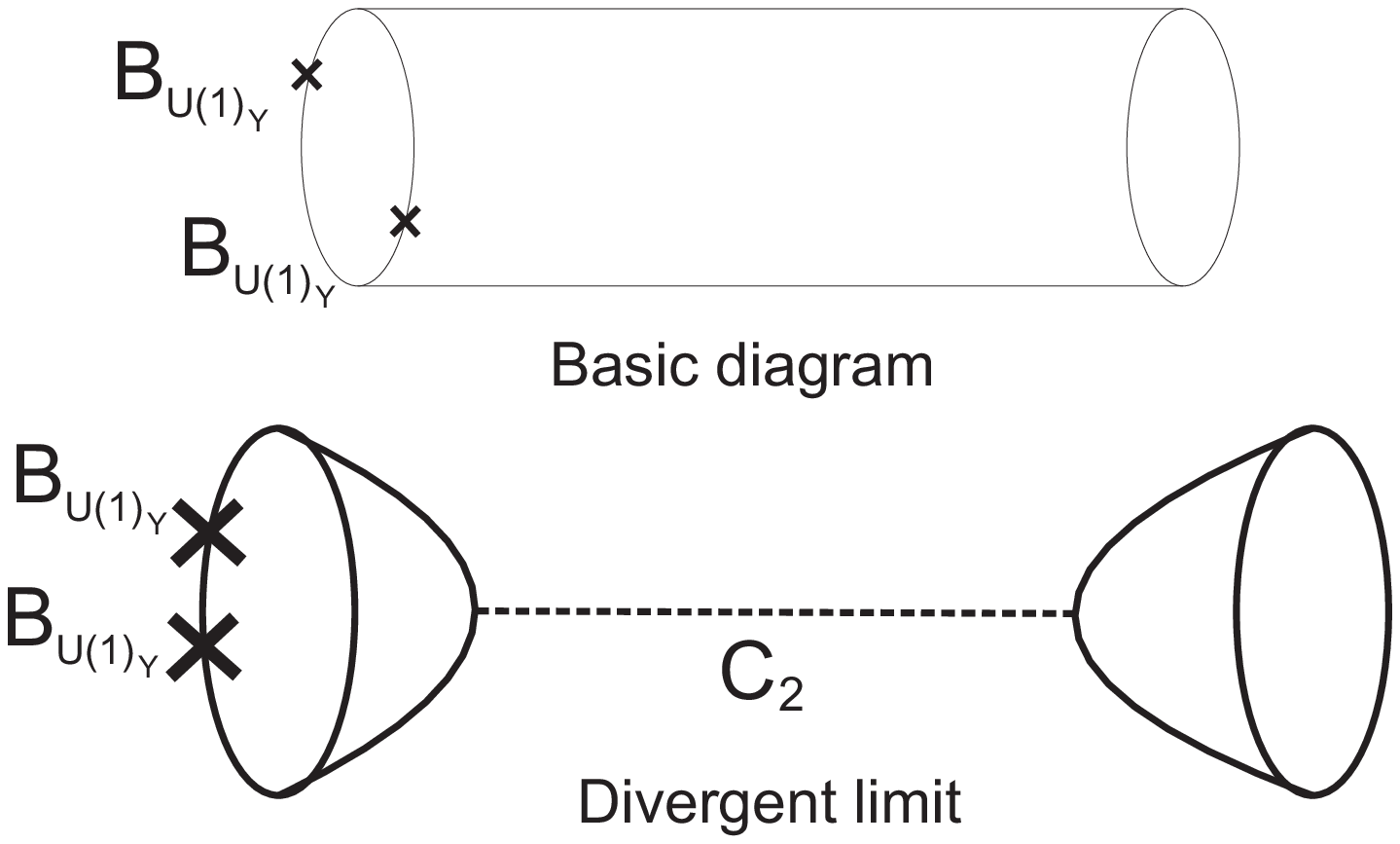}
\caption{\label{fig:epsart} The locally divergent string diagram in the presence of hypercharge flux breaking.
The diagram consists of a 1-loop open string diagram in the presence of a background spacetime gauge field (or equivalently
with the insertion of two gauge vertex operators).
The ultraviolet limit of this diagram is the infrared limit of the closed string tree-diagram,
which diverges due to the intermediate
massless propagator. }
\end{figure}
In the orbifold case the closed string RR tadpoles could be split into $\mc{N}=1$ fully twisted tadpoles,
that were necessarily cancelled at the singularity, and $\mc{N}=2$ partially twisted tadpoles that
could be cancelled by sources/sinks in the bulk.
1-loop gauge coupling running associated with $\mc{N}=1$ sectors therefore runs until the string
scale where the oscillator modes cut it off, while $\mc{N}=2$ running persisted until
the winding mode scale which probes the bulk.
In the GUT case the analogues of $\mc{N}=1$ fully twisted sectors are the RR fields associated to reduction of
$C_2$ or $C_4$ on either the del Pezzo 4-cycle or its canonical 2-cycle.
The analogues of $\mc{N}=2$ sectors are RR fields coming from reducing on
two-cycles of the del Pezzo whose dual four-cycle is not local.

The string diagram that computes threshold corrections is a 1-loop open string diagram with
the insertion of gauge boson vertex operators for any of the $SU(5)$ gauge fields.
Equivalently, using the background field formalism it is sufficient to compute the 1-loop vacuum energy in the presence of
an infinitesimal vev for the spacetime field strength of the $SU(5)$ gauge fields.
In this diagram, all open strings that are charged under the background $SU(5)$ run in the loop; letting the left-hand
end be charged, the right-hand end can end anywhere.
This diagram is shown in
figure 1 and in a purely local model diverges in the (open string) ultraviolet.

The divergence is most easily seen by regarding the diagram as a closed string tree diagram, where an ultraviolet
open string divergence
would correspond to an infrared closed string divergence associated to massless modes.
The number of such massless modes is set by the number of non-trivial cycles and is counted by homology dimensions,
and the presence of a divergence can be investigated using supergravity techniques.

Now recall that GUT breaking is entirely sourced by the hypercharge flux $\mc{F}_Y$, which is
necessarily valued
on a globally trivial but locally non-trivial cycle (i.e. GUT breaking is associated to an $\mc{N}=2$ sector).
$\mc{F}_Y$ can potentially induce a closed string tadpole via the Chern-Simons term
\be
\int_{\mbb{M}^4} \tilde{C}^{\alpha}_0 F_{U(1)_Y}  \wedge F_{U(1)_Y} \int_{D} {\cal F}_Y \wedge \omega_{\alpha}    \;, \label{tadpole}
\ee
 where we reduce $C_2 = \tilde{C}_0^{\alpha} \wedge \omega_{\alpha}$ (and likewise for $SU(3)$ and $SU(2)$).
As $\mc{F}_Y$ has a vacuum vev and $F_{U(1)_Y}$ obtains an effective vev in the background field formalism,
this represents a tadpole for the field $C_2$.
In a purely local model, this tadpole induces a (logarithmic) divergence associated to the propagator for the
massless field $\tilde{C}_0^{\alpha}$.

As for the Green-Schwarz term (\ref{GS}) this interaction actually vanishes for all $\tilde{C}^{\alpha}_0$ due to the global triviality
 of $\mc{F}_Y$.
 However the key point is that this global triviality requires bulk knowledge and any purely local computation is insensitive
 to this feature. In closed string channel, we can regard the
 field $\tilde{C}^{\alpha}_0$ as propagating into the bulk until it becomes sensitive
 to global geometry. Only at this point is the tadpole revealed to be trivial and the divergence regulated.
 Viewing the same string diagram from open string channel,
 this implies that the 1-loop gauge couplings continue to diverge logarithmically until
 winding strings are included in the diagram.

Such winding strings have masses $M \sim R M_s$, which
 is precisely the scale that arose in the field theory analysis.
The global triviality of the cycle on which the hypercharge flux is valued is an
  essential ingredient in F-theory GUTs. As a consequence,
the divergence of running gauge couplings until the scale $R M_s$
  will be a generic feature of all GUT models
 utilising hypercharge flux breaking.

\begin{figure}
\includegraphics[height=4.5cm]{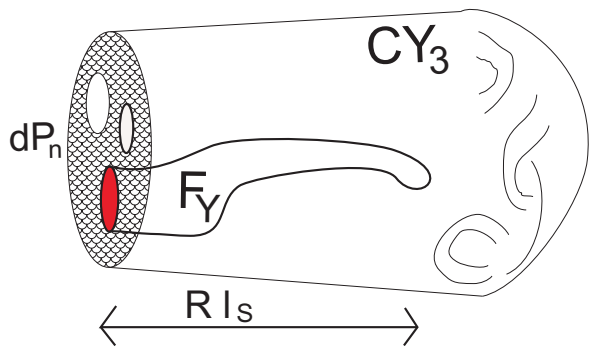}
\caption{\label{fig:epsart} The schematic geometry. The hypercharged flux is on a cycle that is nontrivial locally and trivial globally.}
\end{figure}

\section{GUTs and warped throats}

In the previous sections we have argued that gauge coupling unification occurs at the winding scale because the running continues until the bulk is 
reached. Geometrically this arose because running continued until the cycle wrapped by the hypercharge flux is revealed to be globally trivial.

This physics will also apply in other situations. For example,
 consider a local GUT with hypercharge breaking down a warped throat. As previously we assume that $\mc{F}_Y$ is non-trivial in
 $H^2\left(D, \mbb{Z} \right)$ but is trivial in $H^2\left({\cal M}_6, \mbb{Z} \right)$. Similarly we assume that global triviality requires the cycle to extend out of 
the throat
 and into the bulk region. From a closed string perspective this implies that the $\tilde{C}_0$ fields propagate into the bulk and tadpole
 cancellation does not occur inside the throat.

 From an open string perspective, this implies that gauge coupling running continues until the inclusion of open string modes extending out 
of the throat before returning to the brane. As such winding modes extend into the bulk they have
 masses set by the bulk Planck scale $M_P$ rather than the IR scale of the throat. The cutoff for gauge coupling running is therefore
 $M_P$ rather than the naive value of the local string scale $M_s$ which is warped down.
 This gives the intriguing possibility that a TeV fundamental string scale could be reconciled with unification close to the Planck scale.

While the geometry implies finiteness of threshold corrections will not occur until Planck-scale modes are included in the running,
there may be large corrections to the $\beta$ functions due to the contribution of heavy modes inside the warped throat.
 In contrast to the large volume case it is not possible to constrain the form of these using the Kaplunovsky-Louis formula.
 This would require the precise
 form of the K\"ahler potential and matter metrics for a GUT model in the warped throat. Despite recent progress \cite{Douglas:2008jx} such 
expressions do not yet seem
 available.

\section{Conclusions}

This paper has described how in local GUTs with hypercharge breaking the unification scale is the enhanced scale $R M_s$ rather than the naive
scale $M_s$.
This follows from consistency with the holomorphy properties of supergravity.

For local models this is a significant correction and for unification at $10^{16} \hbox{GeV}$ requires a string scale
at $10^{15} \hbox{GeV}$, a factor of ten smaller than the naive value.
In the context of the LARGE volume models \cite{hepth0502058} this factor may be important for combining TeV-scale soft terms with
gauge unification at the GUT scale \cite{09063297}.

The presence of the bulk radius $R$ in the unification scale also
shows that it is not possible to completely decouple the bulk in phenomenological analyses of the local model.

{\noindent \bf Acknowledgements}

We thank Fernando Quevedo, Stuart Raby and Graham Ross for useful discussions and explanations. JC is supported by a Royal Society University Research 
Fellowship. EP is supported by a STFC Postdoctoral Fellowship.

%
%

\end{document}